\documentclass[twocolumn,showpacs,epsfig,pre]{revtex4}

\usepackage{graphicx}
\usepackage{amssymb}
\usepackage{color}

\newcommand{\figsizeone}{0.48}
\newcommand{\figsizetwo}{0.8}

\begin{document}

\draft
\title{Antiresonance induced by symmetry-broken contacts in quasi-one-dimensional lattices}
\author{Jung-Wan Ryu}
\author{Nojoon Myoung}
\author{Hee Chul Park}
\email{hcpark@ibs.re.kr}
\affiliation{Center for Theoretical Physics of Complex Systems, Institute for Basic Science (IBS), Daejeon 34051, Republic of Korea}
\date{\today}

\begin{abstract}

We report the effect of symmetry-broken contacts on quantum transport in quasi-one-dimensional lattices. In contrast to 1D chains, transport in quasi-one-dimensional lattices, which are made up of a finite number of 1D chain layers, is strongly influenced by contacts. Contact symmetry depends on whether the contacts maintain or break the parity symmetry between the layers. With balanced on-site potential, a flat band can be detected by asymmetric contacts, but not by symmetric contacts. In the case of asymmetric contacts with imbalanced on-site potential, transmission is suppressed at certain energies. We elucidate these energies of transmission suppression related to antiresonance using reduced lattice models and Feynman paths. These results provide a nondestructive measurement of flat band energy which it is difficult to detect.
\end{abstract}
\maketitle
\narrowtext

\section{Introduction}

Quantum dots are often referred to as artificial atoms, as their electrons are confined in all spatial dimensions with discretized energy \cite{Kas93, Ash96}. Likewise, coupled quantum dots can be considered as artificial molecules or crystals \cite{Hol01, Hol02, Sha01}. Quantum dot arrays are accordingly a promising candidate for artificial lattices, which have attracted great attention recently considering their relation to not only fundamental physics but also various potential applications, e.g., quantum computation, spintronics, engineering of energy bands, and topological states in non-Hermitian lattices \cite{Bra13, Hen17, Nak17, Zhe15, Bab16, Cer16, Esa11, Lee16}. The electronic states in ideal arrays of uniform quantum dots can be described by discrete-level representations, such as the tight-binding model \cite{Dat97, Ryn16}, and electronic transport in quantum dot arrays has been extensively studied as the electronic states can be probed when coupling is allowed between dots and leads. In the presence of a defect, transport shows antiresonance where the transmission probability vanishes \cite{Wan02, Ore03, Ore03b, Bao05}; in other words, the conducting channel is completely blocked by the defect when the Fermi energy is right at the defect energy level.

Apart from a one-dimensional chain of quantum dots, quasi-one-dimensional (quasi-1D) lattices have revealed peculiar electronic states, such as flat bands. Flat band lattice models where at least one band is completely flat over the whole momentum space---implying zero dispersion---have attracted considerable interest from various areas including superconductors \cite{Sim97, Den98, Ima00, Den03}, optical and photonic lattices \cite{Apa10, Hyr13, Rec13, Vic15, Muk15}, and exciton-polariton condensates \cite{Jac14, Bab16}. The simplest flat band lattice models are quasi-1D lattices that have translational symmetry in the infinite longitudinal direction and parity symmetry in a finite transverse direction, for example cross-stitch and tunable diamond lattices. In these lattices, the flat and dispersive bands are completely decoupled because they have odd and even parity symmetric eigenstates, respectively. The odd symmetry of the flat bands produces compact localized eigenstates with nonzero amplitude only at a finite number of lattice sites due to destructive interference \cite{Der06, Ber08, Der10, Fla14}.

In this work, we study quantum transport in quasi-1D lattices with symmetric and asymmetric contacts which, in contrast to transport in 1D chains, strongly influence transport in quasi-1D lattices. Imperfect contacts between system and leads can be considered as defects which break the symmetries of the lattice. We consider two cases according to the existence of flat bands related to the balance of on-site potential. First, when a quasi-1D lattice has flat bands due to balanced on-site potential, it is natural that transport does not reflect flat bands as the compact localized states in the flat bands are such that they are unavailable for charge transfer. We show, however, that flat bands can be detected by symmetry-broken contacts, but not by symmetric contacts. Next, in quasi-1D lattices with imbalanced on-site potential, although there are no flat bands, transmission has additional dips irrespective of the energy bands in the case of asymmetric contacts. We explain the energies of this transmission suppression related to antiresonance using reduced lattice models and Feynman paths.

This paper is organized as follows. In Section II we describe the energy bands and quantum transport in quasi-1D lattices with flat bands. The results of transmission in lattices without flat bands due to imbalanced on-site potential are presented in Section III, and in Section IV we summarize our results.

\section{Quasi-one-dimensional lattices - Flat band lattices}

\subsection{Cross-stitch lattices}

\begin{figure}
\begin{center}
\includegraphics[width=\figsizeone\textwidth]{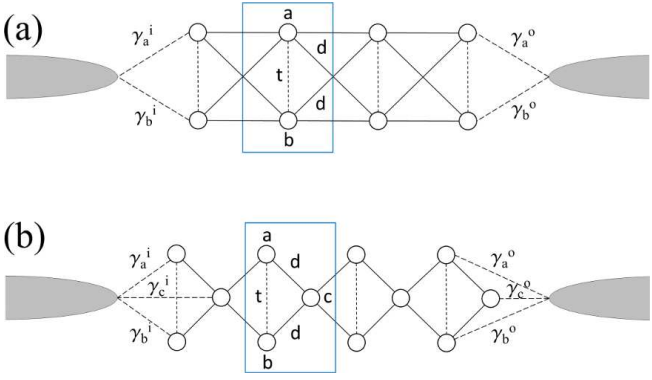}
\caption{(color online).
(a) Cross-stitch lattice with two leads. Blue rectangular boxes represent unit cells.
The unit cell has two sites, $a$ and $b$, with hopping strengths $d$ (solid lines) and $t$ (dashed line) between the sites.
The coupling between sites and leads is $\gamma_{a(b,c)}^{i(o)}$ (long-dashed line).
(b) Tunable diamond lattice with two leads. The unit cell has three sites, $a$, $b$, and $c$, with hopping strengths $d$ (solid lines) and $t$ (dashed line) between sites.
}
\label{fig1}
\end{center}
\end{figure}

\begin{figure*}
\begin{center}
\includegraphics[width=\figsizetwo\textwidth]{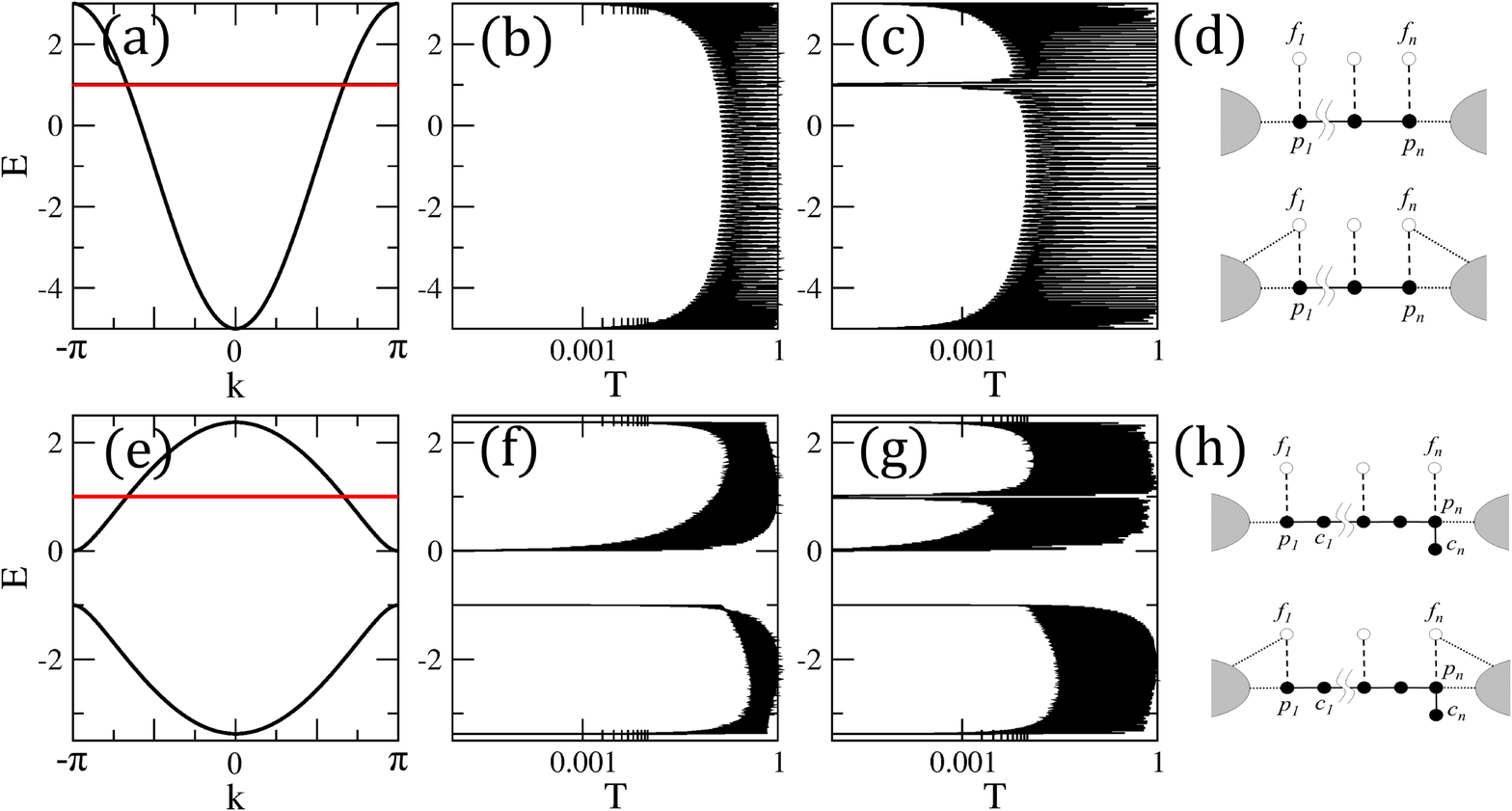}
\caption{(color online).
(a) Energy bands for a cross-stitch lattice with a flat band (red line).
(b) Transmission probabilities for a cross-stitch lattice of which both $a$- and $b$-sites of the end unit cells are connected to the input and output leads.
(c) Transmission probabilities for a cross-stitch lattice of which only the $a$-sites of the end unit cells are connected to the input and output leads.
(d) Upper and lower panels show detangled Fano lattices from a cross-stitch lattice with symmetric and asymmetric contacts, respectively, corresponding to (b) and (c).
The hopping strength (dashed lines) between Fano states $f_n$ and dispersive chain $p_n$ equals zero.
(e) Energy bands for a tunable diamond lattice with a flat band (red line).
(f) Transmission probabilities for a tunable diamond lattice of which both $a$- and $b$-sites of the end unit cells are connected to the input and output leads.
(g) Transmission probabilities for a tunable diamond lattice of which only the $a$-sites of the end unit cells are connected to the input and output leads.
(h) Upper and lower panels show detangled Fano lattices from a tunable diamond lattice with symmetric and asymmetric contacts, respectively, corresponding to (f) and (g).
}
\label{fig2}
\end{center}
\end{figure*}

In this section, we study the energy bands and quantum transport in two quasi-1D lattices: a cross-stitch and a tunable diamond lattice.
Figure~\ref{fig1} illustrates the cross-stitch and tunable diamond lattices. The Schr{\"o}dinger equation of a quasi-1D flat band lattice is given by
\begin{equation}
 E \Psi_j = H_0 \Psi_j + H_1 \Psi_{j+1} + H_1^+ \Psi_{j-1},
\end{equation}
where $H_0$ is on-site energy matrix and $H_1$ is nearest neighbor hopping matrix.
The tight-binding Hamiltonian for a cross-stitch lattice is given by
\begin{equation}
 H_0 = \left(\begin{array}{cc}
 \epsilon_a & -t \\
 -t & \epsilon_b \\
\end{array}\right),
 H_1 = \left(\begin{array}{cc}
 -d & -d \\
 -d & -d \\
\end{array}\right),
\end{equation}
where $\Psi_j = (a_j ~ b_j)^T$. On-site potential energies are $\epsilon_a$ and $\epsilon_b$, respectively, and hopping strengths between the two sites are $d$ and $t$. We can set $\Psi_{j+1} = \Psi_j e^{ik}$ and $\Psi_{j-1} = \Psi_j e^{-ik}$ because of the translational symmetry of the unit cells.
Finally, the Hamiltonian for the cross-stitch lattice is given by
\begin{equation}
 H = \left(\begin{array}{cc}
 \epsilon_a - 2 d \cos{k} & -t - 2 d \cos{k} \\
 -t - 2 d \cos{k} & \epsilon_b - 2 d \cos{k} \\
\end{array}\right).
\end{equation}
Solving the eigenproblem of $H$ when $\epsilon_a = \epsilon_b = 0$, we obtain the energy bands for the cross-stitch lattice as
\begin{equation}
E(k) = -t-4 d \cos{k}, ~E_{FB}=t,
\end{equation}
where $E(k)$ and $E_{FB}$ are dispersive and flat band energies, respectively. Figure~\ref{fig2} (a) shows the dispersive and flat bands when $t=d=1$.

Here, we discuss quantum transport in the quasi-1D lattices.
The system under study is composed of a cross-stitch or tunable diamond lattice with $N$ unit cells, as shown in Fig.~\ref{fig1},
with two leads connected to the left and right end unit cells.
The Hamiltonian of this system is given by
\begin{equation}
 H = H_{Q1D} + H_{lead} + H_{coupling},
\label{eq_trans}
\end{equation}
where $H_{Q1D}$, $H_{lead}$, and $H_{coupling}$ describe the quasi-1D lattice, leads, and coupling between the lattice and leads, respectively, and are given by
\begin{eqnarray}
 H_{Q1D} &=& \sum_{i=1}^{N}H_0 d_{i}^{\dagger} d_{i} + \sum_{i=1}^{N-1} ~(H_{1} d_{i+1}^{\dagger} d_{i} + h.c.) \\
 H_{lead} &=& -\frac{V_0}{2} \sum_{j \neq 0} (c_{j+1}^{\dagger} c_{j} + h.c.) \\
 H_{coupling} &=& - G^i d_{1}^{\dagger} c_{-1} - G^o d_{N}^{\dagger} c_{1} + h.c.,
\end{eqnarray}
where $d_{j}^{\dagger}$ ($d_j$) and $c_{j}^{\dagger}$ ($c_j$) are electron creation (annihilation) operators for the lattice and leads, respectively. $V_0 / 2$ is hopping strength in the leads and $G^{i(o)}$ describes the coupling between the lattice and the left (right) lead. The amplitude equations of the total Hamiltonian of Eq. (\ref{eq_trans}) can be written as
\begin{eqnarray}
 E \phi_{-1} &=& -\frac{V_0}{2} \phi_{-2} + {G^i}^T  \Psi_1\\
 \label{leftc}
 E \Psi_{1} &=& H_{0} \Psi_{1} + H_{1} \Psi_{2} + \phi_{-1} {G^i} \\
 E \Psi_{j} &=& H_{0} \Psi_{j} + H_{1}^{\dagger} \Psi_{j-1} + H_{1} \Psi_{j+1} (2 \leq j \leq N-1) \\
 \label{rightc}
 E \Psi_{N} &=& H_{0} \Psi_{N} + H_{1}^{\dagger} \Psi_{N-1} + \phi_{1} {G^o}\\
 E \phi_{1} &=& -\frac{V_0}{2} \phi_{2} + {G^o}^T \Psi_N 
\end{eqnarray}
where
\begin{eqnarray}
 \phi_{j} =& e^{i q j} + r e^{-i q j} & (j < 0) \\
          =& t e^{i q j} & (j > 0). 
\end{eqnarray}
$\phi_{j}$ represents the $j$th sites of leads and $G^{i(o)}$ is given by
\begin{equation}
\label{geq}
G^{i(o)} = \left(\begin{array}{c}
 -\gamma_a^{i(o)} \\
 -\gamma_b^{i(o)} 
\end{array}\right).
\end{equation}
$r$ and $t$ are reflection and transmission coefficients, respectively, and $\left|r\right|^2 + \left|t\right|^2 = 1$ in Hermitian cases.
Finally, we obtain the equations as follows
\begin{eqnarray}
 -\frac{V_0}{2} &=& \frac{V_0}{2} r + {G^i}^T \Psi_1 \\
 - e^{-i q} {G^i} &=& e^{i q} r {G^i} + (H_{0} - E) \Psi_{1} + H_{1} \Psi_{2} \\
 0 &=& H_{1}^{\dagger} \Psi_{j-1} + (H_{0} - E) \Psi_{j} + H_{1} \Psi_{j+1} \\
 0 &=& H_{1}^{\dagger} \Psi_{N-1} + (H_{0} - E) \Psi_{N} + e^{i q} t {G^o} \\
 0 &=& \frac{V_0}{2} t + {G^o}^T \Psi_N
\end{eqnarray}
where the energy of the leads is given by $e^{\pm i q} = - E / V_{0} \pm i \sqrt{1 - \left|E / V_{0}\right|^2}$.
Finally, we can obtain $R$ and $T$ for the cross-stitch lattice from the following equation
\begin{widetext}
\begin{equation}
 \left(\begin{array}{c}
 -\frac{V_0}{2} \\
 - e^{-iq} G^{i} \\
 0 \\
 \vdots \\
 0 \\
 0 \\
 0 \\
\end{array}\right)=
 \left(\begin{array}{ccccccc}
 \frac{V_0}{2} & {G^{i}}^T & & & & & \\
  e^{iq} G^{i} & H_0 - E I & H_1 & & & &  \\
  & H_1^{\dagger} & H_0 - E I & H_1 & & &  \\
  & & \ddots & \ddots & \ddots && \\
  &&& H_1^{\dagger} & H_0 - E I & H_1 & \\
  &&&& H_1^{\dagger} & H_0 - E I & e^{iq} G^{o}\\
  &&&&& {G^{o}}^T & \frac{V_0}{2} \\
\end{array}\right)
\left(\begin{array}{c}
 r \\
 \Psi_1 \\
 \Psi_2 \\
 \vdots \\
 \Psi_{N-1} \\
 \Psi_{N} \\
 t \\
\end{array}\right).
\end{equation}
\end{widetext}
Hamiltonians $H_0$ and $H_1$ are $2 \times 2$ matrices which describe the unit cell and the coupling between nearest unit cells, respectively. 
We set $V_0 = 10$ throughout this paper.

Figure~\ref{fig2} (b) shows transmission probability $T=\left|t\right|^2$ as a function of energy $E$ in a cross-stitch lattice of $100$ unit cells with symmetric contacts---that is, $\gamma_a^{i}=\gamma_b^{i}=\gamma_a^{o}=\gamma_b^{o}=\gamma$, and $t = d = 1$. We set $\gamma=1$ throughout this paper. The transmission probability corresponds to the dispersive energy bands because the compact localized states in the flat band are such that they are unavailable for transmission.
In lattices including both flat and dispersive bands, the compact localized part of the lattice related to flat bands can be detangled from the dispersive part of the lattice using a generic transformation \cite{Fla14}. After the transformation, we obtain the dispersive part of the lattice and side-coupled Fano states, where the hopping strength between them is determined by the difference between the on-site potentials of two sites in a unit cell. As a result, if the on-site potentials are symmetric, there is no hopping between the dispersive part of the lattice and the Fano states; thus, the Fano states completely decouple from the dispersive part of the lattice. In this case, the Fano state is the compact localized state of the flat band. Conversely, an asymmetric potential intertwines the dispersive and Fano states. This detangling procedure clearly shows the relation between the local symmetry of the lattice and the compact localized state, and explains why the flat band state does not contribute to transmission.

Considering a cross-stitch lattice with symmetric contacts, the amplitude equations of the left end unit cell can be written from Eq.~(\ref{leftc}) as
\begin{eqnarray}
E a_1 &=& \epsilon_a a_1 - a_2 - b_2 - b_1 -\gamma \phi_{-1}, \\\nonumber
E b_1 &=& \epsilon_b b_1 - b_2 - a_2 - a_1 -\gamma \phi_{-1},
\end{eqnarray}
where $a_1$ and $b_1$ represent the $a$- and $b$-sites of the left end unit cell and $\phi_{-1}$ represents the first site of the left lead. From these equations, we obtain the end unit cell of a lattice with dispersive degrees of freedom $p_1$ and side-coupled Fano states $f_1$,
\begin{eqnarray}
\label{fanoeq}
E p_1&=& (\epsilon^{+} - 1) p_1 + \epsilon^{-} f_1 - 2 p_2 - \sqrt{2} \gamma \phi_{-1}, \\\nonumber
E f_1 &=&  (\epsilon^{+} + 1) f_1 + \epsilon^{-} p_1,
\end{eqnarray}
where $p_n = (a_n + b_n)/\sqrt{2}$, $f_n = (a_n - b_n)/\sqrt{2}$, $\epsilon^{+} = (\epsilon_{a} + \epsilon_{b})/2$, and $\epsilon^{-} = (\epsilon_{a} - \epsilon_{b})/2$. If $\epsilon^{-} = 0$, the side-coupled Fano state $f_1$ becomes a compact localized state of which energy is $\epsilon^{+} +1$ because it is isolated in spite of the coupling between the lattice and leads. In the detangled lattice in the upper panel of Fig.~\ref{fig2} (d), the Fano states become compact localized states and do not affect transmission because the Fano states are completely decoupled from the dispersive chain, i.e., $\epsilon^{-} = 0$.

However, regarding asymmetric contacts, where $\gamma_a^{i} = \gamma_a^{o} = \gamma$ and $\gamma_b^{i}=\gamma_b^{o}=0$, a sharp dip in the transmission appears at flat band energy $E=1$, as shown in Fig.~\ref{fig2} (c).
This dip in transmission originates from the breaking of local symmetry of the end unit cells due to the asymmetric contacts.
When we consider asymmetric contacts, Eq.~(\ref{fanoeq}) changes into
\begin{eqnarray}
\label{fanoeq_a}
E p_1&=& (\epsilon^{+} - 1) p_1 + \epsilon^{-} f_1 - 2 p_2 - \frac{\gamma}{\sqrt{2}} \phi_{-1}, \\\nonumber
E f_1 &=&  (\epsilon^{+} + 1) f_1 + \epsilon^{-} p_1 - \frac{\gamma}{\sqrt{2}} \phi_{-1}.
\end{eqnarray}
If $\epsilon^{-} = 0$, the $f_n (n=2,\dots,n-1)$ inside the lattice are still compact localized states, except for $f_1$ because of the coupling between $f_{1}$ and the lead as shown in the lower panel of Fig.~\ref{fig2} (d). The coupling between Fano states and leads due to asymmetric contacts produces the dip in transmission at the flat band energy though most compact localized states still exist inside the lattice. Considering the amplitude equations for the right unit cell written from Eq.~(\ref{rightc}), we obtain the same results. It is noted that this transmission dip can be also considered as an antiresonance due to destructive interference, of which position corresponds to the energy of the Fano state \cite{Wan02, Ore03, Ore03b, Bao05}. 

\subsection{Tunable diamond lattices}

In the case of a tunable diamond lattice, the tight-binding Hamiltonian is given by
\begin{equation}
 H_0 = \left(\begin{array}{ccc}
 \epsilon_a & -t & -d \\
 -t & \epsilon_b & -d \\
 -d & -d & \epsilon_c \\
\end{array}\right),
 H_1 = \left(\begin{array}{ccc}
 0 & 0 & 0 \\
 0 & 0 & 0 \\
 -d & -d & 0 \\
\end{array}\right),
\end{equation}
where $\Psi_j = (a_j ~ b_j ~ c_j)^T$. On-site potential energies are $\epsilon_a$, $\epsilon_b$, and $\epsilon_c$, respectively, and hopping strengths between the sites are $d$ and $t$. Setting $\Psi_{j+1} = \Psi_j e^{ik}$ and $\Psi_{j-1} = \Psi_j e^{-ik}$ on account of the translational symmetry of the unit cells,
we obtain the Hamiltonian for the tunable diamond lattice as
\begin{equation}
 H = \left(\begin{array}{ccc}
 \epsilon_a & -t & -d-d e^{-ik} \\
 -t & \epsilon_b & -d-d e^{-ik} \\
 -d-d e^{ik} & -d-d e^{ik} & \epsilon_c
\end{array}\right).
\end{equation}
Solving the eigenproblem of $H$ when $\epsilon_a = \epsilon_b = \epsilon_c = 0$,
we obtain the energy bands for the tunable diamond lattice as follows
\begin{equation}
E_{1,2}(k) = -\frac{1}{2} \left(t \pm \sqrt{t^2 + 32 d^2 \cos^2{\frac{k}{2}}} \right), ~E_{FB}=t,
\end{equation}
where $E_{1,2}(k)$ and $E_{FB}$ are dispersive and flat band energies, respectively. Figure~\ref{fig2} (e) shows the dispersive and flat bands when $t=d=1$.

Considering the transport problem in the tunable diamond lattice, $G^{i(o)}$ in Eq.~(\ref{geq}) changes into 
\begin{equation}
G^{i(o)} = \left(\begin{array}{c}
 -\gamma_a^{i(o)} \\
 -\gamma_b^{i(o)} \\
 -\gamma_c^{i(o)}
\end{array}\right)
\end{equation}
and $H_0$ and $H_1$ are now $3 \times 3$ matrices for the tunable diamond lattice.
Figure~\ref{fig2} (f) shows transmission $T=\left|t\right|^2$ as a function of energy $E$ in a tunable diamond lattice of $100$ unit cells with symmetric contacts, e.g., $\gamma_a^{i}=\gamma_b^{i}=\gamma_a^{o}=\gamma_b^{o}=\gamma$ and $\gamma_c^{i}=\gamma_c^{o}=0$. The transmission probability corresponds to the dispersive energy bands.
In the case of asymmetric contacts, where $\gamma_a^{i} = \gamma_a^{o} = \gamma$ and $\gamma_b^{i}=\gamma_b^{o}=\gamma_c^{i}=\gamma_c^{o}=0$, there is again a dip in transmission at flat band energy $E=1$, as shown in Fig.~\ref{fig2} (g). Transmission in tunable diamond lattices is similar to that in cross-stitch lattices as the detangling of the flat bands is similar, except for the $c$-sites, as shown in Fig.~\ref{fig2} (h).

\section{Quasi-one-dimensional lattices with imbalanced on-site potential}

In this section, we study energy bands and quantum transport in quasi-1D lattices with imbalanced on-site potential.
While there are no flat bands in this case, as the symmetry for flat bands is broken by the imbalanced on-site potential, here we are able to find bandgaps.
We now consider quantum transport in quasi-1D lattices with imbalanced on-site potential. In our case, there is no flat band as the imbalanced potential breaks the flat band and opens up bandgaps, as shown in Fig.~\ref{fig3} (a) and Fig.~\ref{fig4} (a).

\subsection{Cross-stitch lattices}

\begin{figure}
\begin{center}
\includegraphics[width=\figsizeone\textwidth]{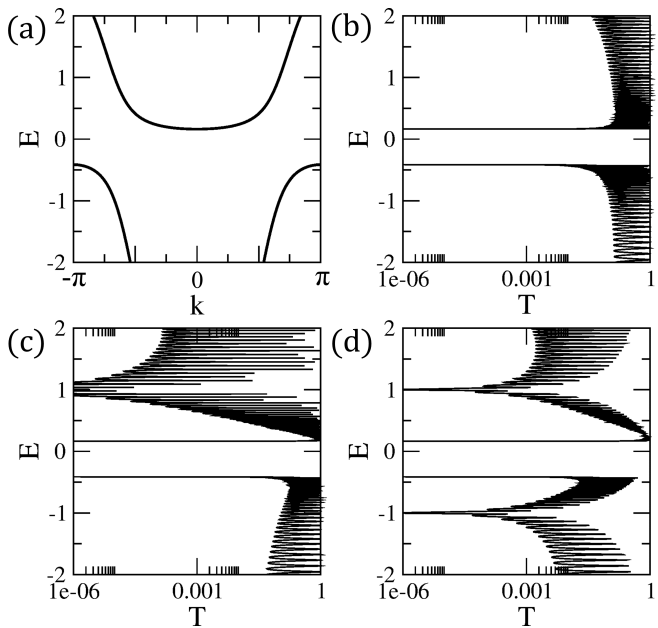}
\caption{(color online).
(a) Energy bands for a cross-stitch lattice with $\delta_a = -2$.
(b) Transmission probability for a cross-stitch lattice of which both $a$- and $b$-sites of the end unit cells are connected to the input and output leads.
(c) Transmission probability for a cross-stitch lattice of which only the $a$-sites of the end unit cells are connected to the input and output leads.
(d) Transmission probability for a cross-stitch lattice of which $a$- and $b$-sites of the end unit cells are connected to the input and the output leads, respectively.
}
\label{fig3}
\end{center}
\end{figure}

\begin{figure}
\begin{center}
\includegraphics[width=\figsizeone\textwidth]{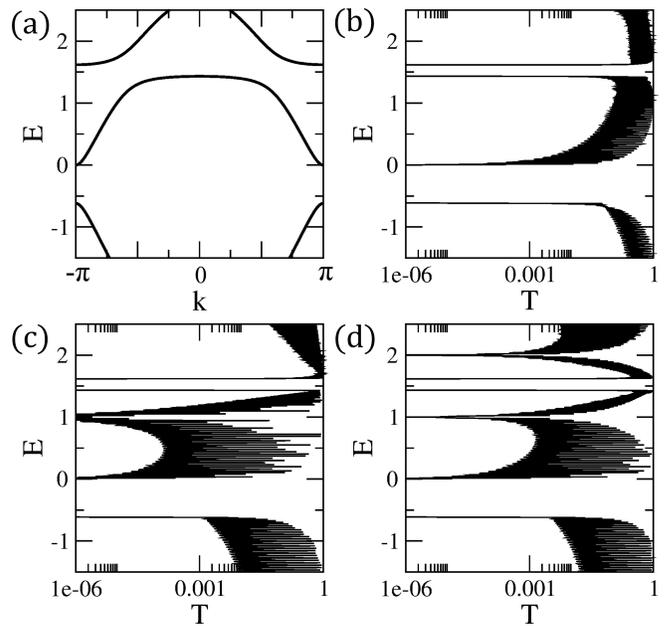}
\caption{(color online).
(a) Energy bands for a tunable diamond lattice with $\delta_a = 1$.
(b) Transmission probability for a tunable diamond lattice of which both $a$- and $b$-sites of the end unit cells are connected to the input and output leads.
(c) Transmission probability for a tunable diamond lattice of which only the $a$-sites of the end unit cells are connected to the input and output leads.
(d) Transmission probability for a tunable diamond lattice of which $a$- and $b$-sites of the end unit cells are connected to the input and the output leads, respectively.
}
\label{fig4}
\end{center}
\end{figure}

In order to study the effect of imbalanced on-site potential on a cross-stitch lattice, we apply on-site potentials to the diagonal parts of $H_0$ as follows
\begin{equation}
 H_0 = \left(\begin{array}{cc}
 \epsilon_0 + \delta_a & -t \\
 -t & \epsilon_0 + \delta_b \\
\end{array}\right),
\end{equation}
where $\delta_a$ and $\delta_b$ are imbalanced on-site potential, or in other words $\delta_a \neq \delta_b$.
If we apply imbalanced on-site potential, e.g. $\delta_a = -2$,
the two bands exhibit a repulsive behavior that results in a band gap, as seen in Fig.~\ref{fig3} (a).
In this case, the flat band disappears because the related symmetry is broken.

Figure~\ref{fig3} (b)-(d) shows the transmission probability $T$ when $\delta_a = -2$ and $\epsilon_0 = \delta_b = 0$ in a cross-stitch lattice with different contact configurations. First, we consider that both $a$- and $b$-sites of the end unit cells are connected with the leads. In other words, incident waves from the left lead are transmitted through both $a$- and $b$-sites of the left end unit cell of the lattice to the right lead through both $a$- and $b$-sites of the right end unit cell. In this case, the transmission corresponds with the energy band (cf. Fig.~\ref{fig3} (b)). Second, we consider the case of symmetry-broken contacts, where incident waves from the left lead are transmitted through the left $a$-site of the lattice to the right lead through the right $a$-site. In this case, there exists a dip in transmission at $E=1$, which is irrelevant to the energy bands of the cross-stitch lattice as well as the energy bandgap (cf. Fig.~\ref{fig3} (c)). Finally, we consider the case of different contacts, where incident waves from the left lead are transmitted through the left $a$-site of the lattice to the right lead through the right $b$-site. Here, two dips in transmission appear at $E=1$ and $E=-1$ (cf. Fig.~\ref{fig3} (d)). The last two cases of symmetry-broken contacts show additional dips in transmissions which are irrelevant to the energy bands.

\begin{figure}
\begin{center}
\includegraphics[width=\figsizeone\textwidth]{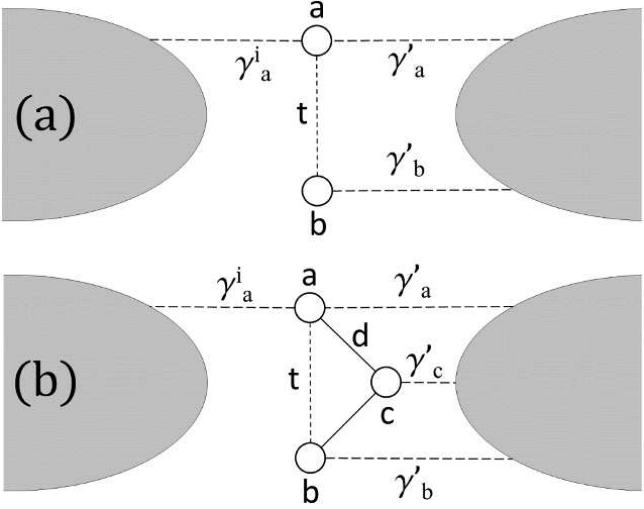}
\caption{(color online).
(a) Schematic diagram of the end unit cells of a cross-stitch lattice connected to the leads.
Both $a$- and $b$-sites are connected to the right lead but only the $a$-site is connected to the left lead.
(b) Schematic diagram of the end unit cells of a tunable diamond lattice connected to the leads.
All sites are connected to the right lead but only the $a$-site is connected to the left lead.
}
\label{fig5}
\end{center}
\end{figure}

In order to understand the dips in transmission due to symmetry-broken contacts in quasi-1D lattices, which cannot be explained by the energy bands of the lattices, let's reconsider Eq.~(\ref{eq_trans}) in detail. The $H_{Q1D}$ term describing the quasi-1D lattices can be divided into two parts, $H'_{Q1D}$ and $H_{end}$, which represent the flat band excluding the two end unit cells and the two end unit cells, respectively.
In this case $H'_{Q1D}$ precisely describes the band and gap structures of the flat band lattices, as our system is sufficiently large.
Therefore, it is $H_{end}$ that produces the additional dips in transmission on account of the asymmetric contacts between the end unit cells and leads.
Figure~\ref{fig5} (a) depicts a reduced cross-stitch lattice with the asymmetric contacts between leads.
When the left lead is connected to the $a$-site only and the right lead is connected to both $a$- and $b$-sites, we can obtain $T$ from the following equation
\begin{equation}
 \left(\begin{array}{c}
 -\frac{V_0}{2} \\
 \gamma_a^{i} e^{-i q} \\
 0 \\
 0 \\
\end{array}\right)=
 \left(\begin{array}{cccc} 	
 \frac{V_0}{2} & -\gamma_a^{i} & 0 & 0 \\
 -\gamma_a^{i} e^{i q} & \epsilon_a - E & -t & -\gamma'_a e^{i q}  \\
 0 & -t & \epsilon_b - E & -\gamma'_b e^{i q} \\
 0 & -\gamma'_a & -\gamma'_b & \frac{V_0}{2} \\
\end{array}\right)
\left(\begin{array}{c}
 r \\
 a \\
 b \\
 t \\
\end{array}\right).
\end{equation}
For simplicity, if $\gamma'_a=\gamma'_b=\gamma'$, the condition for transmission $T=0$ is $E_a = \epsilon_b + t$ when the left lead is connected to the $a$-site. Symmetrically, if the left lead is connected to the $b$-site, then the condition is $E_b = \epsilon_a + t$.
Finally, with asymmetric contacts, the transmission in Fig.~\ref{fig3} (c) and (d) dips at $E=E_a$ when the leads are connected to the $a$-site of the end unit cell and at $E=E_b$ for a $b$-site connection, in addition to the bandgaps of the cross-stitch lattice.

These dips in transmission can be considered as antiresonance in terms of the concept of Feynman path \cite{Dat97, Bao05}.
We can obtain the retarded Green's function $G^r_a$ using the Feynman path for electron transmission through the reduced lattice in Fig.~\ref{fig5} (a) as
\begin{eqnarray}
\label{green}
G^r_a &=& g^r_a - g^r_a t g^r_b + g^r_a t g^r_b t g^r_a - g^r_a t g^r_b t g^r_b t g^r_a + \cdots \\
           &=& \frac{g^r_a (1-t g^r_b)}{1-t^2 g^r_a g^r_b},
\end{eqnarray}
where $g^r_a = 1/(E - \epsilon_a + 2 i \Gamma_0)$ and $g^r_b = 1/(E-\epsilon_b + i \Gamma_0$). The transmission function is associated with Green's function $G^r_a (E)$ by the relation $T(E) \propto \left| G^r_a (E) \right|^2$. We can also obtain the condition, $E = E_a = \epsilon_b+t$, where the interference of all Feynman paths leads to antiresonance when the left lead is connected to the $a$-site and $\Gamma_0$ is infinitesimal.

\subsection{Tunable diamond lattices}

Now, to study the effect of imbalanced on-site potential on a tunable diamond lattice, we apply on-site potentials to the diagonal parts of $H_0$ as follows 
\begin{equation}
 H_0 = \left(\begin{array}{ccc}
 \epsilon_0 + \delta_a & -t & -d \\
 -t & \epsilon_0 + \delta_b & -d \\
 -d & -d & \epsilon_0 + \delta_c \\
\end{array}\right),
\end{equation}
where $\delta_a$, $\delta_b$, and $\delta_c$ are imbalanced on-site potential.
Under symmetric conditions, e.g. when $\delta_c$ is a non-zero real value and $\delta_a=\delta_b=0$, two dispersive bands show repulsive behavior but the flat band survives. As for asymmetric conditions, e.g. when $\delta_a \neq \delta_b$ and $\delta_c = 0$, the flat and the upper dispersive bands demonstrate a repulsive behavior, as shown in Fig.~\ref{fig4} (a), because the imbalanced on-site potential breaks the symmetry between the $a$- and $b$-sites, which is the origin of the flat band in the previous section.

Figure~\ref{fig4} (b)-(d) shows the transmission probability $T$ when $\delta_a = 1$ and $\epsilon_0 = \delta_b = \delta_c = 0$ in a tunable diamond lattice with different contact configurations. 
This case as well presents additional dips in transmission, which can be also understood in the same manner as the former case.
When the left lead is connected to the $a$-site only and the right lead is connected to the $a$-, $b$-, and $c$-sites simultaneously,
as shown in Fig.~\ref{fig5} (b), we can obtain $T$ from the following equation
\begin{equation}
 \left(\begin{array}{c}
 -\frac{V_0}{2} \\
 \gamma_a^{i} e^{-i q} \\
 0 \\
 0 \\
 0 \\
\end{array}\right)=
M
\left(\begin{array}{c}
 r \\
 a \\
 b \\
 c\\
 t \\
\end{array}\right),
\end{equation}
where
\begin{equation}
M=
 \left(\begin{array}{ccccc} 	
 \frac{V_0}{2} & -\gamma_a^{i} & 0 & 0 & 0 \\
 -\gamma_a^{i} e^{i q} & \epsilon_a - E & -t & -d & -\gamma'_a e^{i q}  \\
 0 & -t & \epsilon_b - E & -d & -\gamma'_b e^{i q} \\
 0 & -d & -d & \epsilon_c - E & -\gamma'_c e^{i q} \\
 0 & -\gamma'_a & -\gamma'_b & -\gamma'_c & \frac{V_0}{2} \\
\end{array}\right).
\end{equation}
For simplicity, if $\gamma_a^{'}=\gamma_b^{'}=\gamma_c^{'}=\gamma^{'}$, the condition for transmission $T=0$ is $E_a = \epsilon_b + t$ or $E_a=\epsilon_c + d$ when the left lead is connected to the $a$-site.
As before, a $b$-site connection gives the condition $E_b = \epsilon_a+t$ or $E_b=\epsilon_c + d$.
Finally, with asymmetric contacts, the transmission in Fig.~\ref{fig4} (c) and (d) dips at $E=E_a$ and $E=E_b$ when the lead is connected to the $a$-site and $b$-site of the end unit cell, respectively.
These dips in transmission can also be considered as antiresonance in terms of the Feynman path;
the former condition, $E_a = \epsilon_b + t$, can be obtained from the same Feynman paths as those in the cross-stitch lattice if we consider the Feynman paths including only $a$- and $b$-sites, exclusive of the $c$-site, in Fig.~\ref{fig5} (b). Similarly, the latter condition $E_a=\epsilon_c + d$ can be obtained from the Feynman paths which include $a$- and $c$-sites but exclude the $b$-site. As seen in Fig.~\ref{fig4}, there are multiple antiresonances in the tunable diamond lattice.

\section{Discussion and summary}

In principle, our results demonstrating that the contacts between a system and leads can induce various resonant phenomena can be applied to quantum transport in diverse situations, e.g., in the presence of defects and flat bands \cite{Wan02, Ore03, Ore03b, Bao05, Lop14}. If the contact does not break the intrinsic properties of the system, such as the parity symmetry of a quasi-1D lattice, then the contact does not influence quantum transport. In the case of a symmetry-broken contact, however, antiresonance indirectly reflects non-mobile states as compact localized states, since the contact perturbs one of the flat-band states at the edge of the system. As a result, contact configurations which break the symmetries that determine the intrinsic properties of a quantum state play an important role in transport.

In summary, we have studied quantum transport in quasi-one-dimensional lattices with symmetric and asymmetric contacts. In a quasi-1D lattice with balanced on-site potential, flat bands can be detected through asymmetric rather than symmetric contacts, and in a lattice with imbalanced on-site potential, transmission was suppressed at certain energies in the case of asymmetric contacts. We have elucidated the energies of the transmission suppressions related to antiresonance using reduced lattice models and Feynman paths. 
There is no overlap between compact localized states of a flat band because the states have nonzero amplitude only at a finite number of lattice sites. Consequently, it is very difficult to detect flat band energy in quantum transport because compact localized states do not contribute to the transport. Due to the isolation of compact localized states, however, asymmetric contacts facilitate a nondestructive measurement of flat band energy.

\section*{Acknowledgments}
This work was supported by the Institute for Basic Science in Korea (IBS-R024-D1).

\end{document}